\documentclass[12pt,twoside,a4paper]{article}
\usepackage{graphicx}
\usepackage{url}
\usepackage{color}

\def\alt{\:\raisebox{-0.5ex}{$\stackrel{\textstyle<}{\sim}$}\:}
\def\agt{\:\raisebox{-0.5ex}{$\stackrel{\textstyle>}{\sim}$}\:}

\catcode`@=11
\newcount\@tempcntc
\def\@citex[#1]#2{\if@filesw\immediate\write\@auxout{\string\citation{#2}}\fi
  \@tempcnta\z@\@tempcntb\m@ne\def\@citea{}\@cite{\@for\@citeb:=#2\do
    {\@ifundefined
       {b@\@citeb}{\@citeo\@tempcntb\m@ne\@citea\def\@citea{,}{\bf
?}\@warning
       {Citation `\@citeb' on page \thepage \space undefined}}%
    {\setbox\z@\hbox{\global\@tempcntc0\csname b@\@citeb\endcsname\relax}%
     \ifnum\@tempcntc=\z@ \@citeo\@tempcntb\m@ne
       \@citea\def\@citea{,}\hbox{\csname b@\@citeb\endcsname}%
     \else
      \advance\@tempcntb\@ne
      \ifnum\@tempcntb=\@tempcntc
      \else\advance\@tempcntb\m@ne\@citeo
      \@tempcnta\@tempcntc\@tempcntb\@tempcntc\fi\fi}}\@citeo}{#1}}
\def\@citeo{\ifnum\@tempcnta>\@tempcntb\else\@citea\def\@citea{,}%
  \ifnum\@tempcnta=\@tempcntb\the\@tempcnta\else
   {\advance\@tempcnta\@ne\ifnum\@tempcnta=\@tempcntb \else
\def\@citea{--}\fi
    \advance\@tempcnta\m@ne\the\@tempcnta\@citea\the\@tempcntb}\fi\fi}
\catcode`@=12
\begin{document}
\thispagestyle{empty} 
\title{
\vskip-3cm
{\baselineskip14pt
\centerline{\normalsize DESY 14--186 \hfill ISSN 0418--9833}
\centerline{\normalsize MITP/14--073 \hfill} 
\centerline{\normalsize LPSC-14--260 \hfill} 
\centerline{\normalsize January 2015 \hfill}
} 
\vskip1.5cm
\boldmath
{\bf Inclusive $B$-meson production at small $p_T$ in}
\\
{\bf the general-mass variable-flavor-number scheme}
\unboldmath
\author{
B.~A.~Kniehl$^1$, 
G.~Kramer$^1$, 
I.~Schienbein$^2$ 
and H.~Spiesberger$^3$
\vspace{2mm} \\
\normalsize{
  $^1$ II. Institut f\"ur Theoretische
  Physik, Universit\"at Hamburg,
}\\ 
\normalsize{
  Luruper Chaussee 149, 22761 Hamburg, Germany
} \vspace{2mm}\\
\normalsize{
  $^2$ LPSC, Universit\'e Grenoble-Alpes, CNRS/IN2P3,
}\\
\normalsize{
  53 avenue des Martyrs, 38026 Grenoble, France
} \vspace{2mm}\\
\normalsize{
  $^3$ PRISMA Cluster of Excellence, Institut f\"ur Physik,
}\\
\normalsize{
  Johannes Gutenberg-Universit\"at, 55099 Mainz, Germany,
}\\ 
\normalsize{
  and Centre for Theoretical and Mathematical Physics and 
  Department of Physics,
}\\
\normalsize{ 
  University of Cape Town, Rondebosch 7700, South Africa
} \vspace{8mm} \\
}}

\date{}
\maketitle
\begin{abstract}
\medskip
\noindent
We calculate the cross section for the inclusive production of $B$ mesons in
$pp$ and $p\bar{p}$ collisions at next-to-leading order in the
general-mass variable-flavor-number scheme 
and show that a suitable choice of factorization scales leads 
to a smooth transition to the fixed-flavor-number scheme. Our 
numerical results are in good agreement with data from the 
Tevatron and LHC experiments at small and at large transverse 
momenta.
\\[8mm]
PACS: 12.38.Bx, 12.39.St, 13.85.Ni, 14.40.Nd
\end{abstract}

\clearpage


\section{Introduction}

In the last twenty-five years, there has been much interest in the 
study of $B$-meson production in $p\bar{p}$ and $pp$ collisions 
at hadron colliders, both experimentally and theoretically. 
First measurements were done in the late eighties by the UA1 
Collaboration at the CERN $Sp\bar{p}S$ collider \cite{Albajar:1988th} 
operating at a center-of-mass energy of $\sqrt{S}=0.63$~TeV. Later 
measurements were performed by the CDF and D0 collaborations at the 
Fermilab Tevatron in $p\bar p$ collisions at $\sqrt{S}=1.8$~TeV
\cite{Abe:1995dv,Acosta:2002pw,Abachi:1994kj,Abbott:1999wu}
and at $\sqrt{S}=1.96$~TeV
\cite{Acosta:2004yw,Abulencia:2006ps,Aaltonen:2009xn}. 
Recently, the CMS
\cite{Khachatryan:2011mk,Chatrchyan:2011pw,Chatrchyan:2011vh}, ATLAS
\cite{Aad:2012jga,ATLAS:2013cia}, and 
LHCb \cite{Aaij:2012jd,Aaij:2013noa} collaborations at the CERN 
Large Hadron Collider (LHC) published their results for inclusive 
$B^+$, $B^0$, and $B_s$ meson production in $pp$ collisions at 
$\sqrt{S}=7$~TeV. In all these measurements, the differential 
cross sections $d\sigma/dp_T$, where $p_T$ is the transverse 
momentum of the $B$ meson, integrated over specific rapidity 
($y$) regions, or $d\sigma/dy$ integrated over $p_T \geq p_{T, {\rm min}}$
were obtained. While the 
CMS and ATLAS data were obtained at central rapidities, 
LHCb performed measurements at forward rapidities, 
$2 < y < 4.5$. Actually, only few measurements were able to 
explore the small-$p_T$ range down to $p_{T, {\rm min}} \simeq 0$, 
namely those by CDF \cite{Acosta:2004yw} and the two by
LHCb \cite{Aaij:2012jd,Aaij:2013noa}. In all other data, the cut
$p_{T, {\rm min}} > 5$~GeV was imposed. A unified theoretical description 
of these data, which covers both the very small and the large 
$p_T$ range, requires special efforts, which we shall undertake 
in this paper.

On the theoretical side, it is generally accepted that for the 
treatment of $B$-meson production at small $p_T$ values, as well as 
for the calculation of the integrated cross section including 
the small-$p_T$ range, i.e., with $p_{T}$ of order $O(m_b)$ and 
below, where $m_b$ is the $b$-quark mass, one should use the 
so-called massive scheme or fixed-flavor-number scheme (FFNS) 
\cite{Nason:1987xz,Nason:1989zy,Beenakker:1988bq,Beenakker:1990maa,
Bojak:2001fx}, in which the number of active quark flavors 
in the initial state is limited to $n_f=4$, and the $b$ quark 
appears with explicit mass dependence only in the final state. 
In this case, the $b$ quark is treated as a heavy particle in 
the final state and not as a parton in the initial (anti)proton.

In the large-$p_T$ region, characterized by $p_T \gg m_b$, 
the so-called massless scheme or 
zero-mass variable-flavor-number scheme (ZM-VFNS) 
\cite{Cacciari:1993mq,Kniehl:1995em,Cacciari:1995ej,Binnewies:1997gz,%
Kniehl:1996we,Binnewies:1997xq,Binnewies:1998vm,Kniehl:1999vf,Kniehl:2002xn,%
Kniehl:2005de,Kniehl:2006mw,Cacciari:2012ny}
is considered to be appropriate. This is the conventional parton 
model approach, where the $b$ quark is considered massless like 
any other parton. The $b$ quark is also treated as an incoming 
parton coming from the (anti)proton leading to additional 
contributions from hard-scattering subprocesses besides those 
with $u$, $d$, $s$, and $c$ quarks or the gluon ($g$) in the 
initial state. Although this approach can be used as soon as the 
factorization scales $\mu_I$ and $\mu_F$ associated with the 
initial- and final-state singularities are above the starting 
scale of the parton distribution functions (PDFs) and fragmentation 
functions (FFs), its predictions are reliable only in the large-$p_T$ region,
$p_T \gg m_b$, where terms of the order of 
$m_b^2/p_T^2$ can be safely neglected. A next-to-leading-order 
(NLO) calculation in this scheme automatically resums leading and 
next-to-leading logarithmic (NLL) terms. At the same time, all 
non-logarithmic terms through $O(\alpha_s)$ relative to the 
Born approximation are retained for $m_b = 0$. With the conventional 
choice of renormalization and factorization scales,
$\mu_R = \mu_I = \mu_F = m_T$ with $m_T = \sqrt{p_T^2+m_b^2}$, the results 
are dominated by contributions from the $b$-quark PDF down to 
$p_T \simeq 0$.

On the other hand, the general-mass variable-flavor-number scheme 
(GM-VFNS)
\cite{Kramer:2001gd,Kramer:2003cw,Kramer:2003jw,Kniehl:2004fy,Kniehl:2005mk,%
Kniehl:2005st,Kniehl:2005ej,Kniehl:2008zza,Kneesch:2007ey,Kniehl:2009ar,%
Kniehl:2009mh,Kramer:2009kt,Kniehl:2011bk,Kniehl:2012ti}
provides a theoretical 
framework which combines the FFNS and the ZM-VFNS. The ZM-VFNS 
is extended into the intermediate-$p_T$ range by retaining the 
mass-dependent terms of the FFNS. With the conventional choice of 
scales indicated above, the results in this scheme are also 
dominated by the contributions of the $b$-quark PDF down to 
$p_T \simeq 0$. As a consequence, there is no smooth transition from the 
GM-VFNS at large $p_T$ values to the FFNS in the small-$p_T$ 
range, and the GM-VFNS fails to describe the small-$p_T$ CDF and LHCb 
data. 

The dominance of contributions with $b$-quarks in the initial 
state at small $p_T$ values is linked to the fact that this part is treated 
in the massless scheme, as a calculation of the $b$-quark-initiated 
subprocesses in a scheme with massive partons (like the ACOT scheme
\cite{Aivazis:1993pi}) is not available for hadroproduction.\footnote{ 
  For deep inelastic scattering, heavy-quark-initiated processes at NLO with
 massive 
  quarks have been considered in Ref.~\cite{Kretzer:1998ju}. 
  }
The cross section with massless partons is, however, 
divergent for $p_T \to 0$. For a realistic description, we thus 
have to find a way to eliminate or modify this contribution in 
the small-$p_T$ region. In this paper, we shall develop an 
approach to modify the GM-VFNS in such a way that it matches 
the FFNS with the exact $m_b$ dependence by a suitable choice of 
$\mu_I$ and $\mu_F$. We shall study how these modifications 
can lead to a better agreement with presently available experimental 
data at small $p_T$ values.

The content of this paper is as follows. In Section~2, we introduce 
our strategy for the transition to the FFNS and compare our 
predictions with the CDF \cite{Acosta:2004yw} and the LHCb \cite{Aaij:2013noa}
data. We shall also 
present results to be compared with measurements by the ATLAS 
Collaboration \cite{ATLAS:2013cia}. Our conclusions are presented in 
Section~3.


\boldmath
\section{Small-$p_T$ results and comparisons with data}
\unboldmath

In this section, we shall discuss a viable unified 
framework of theoretical predictions for inclusive $B$-meson 
production at small and large $p_T$ values. We shall compare with the 
cross section distributions $d\sigma/dp_T$ 
measured by CDF \cite{Acosta:2004yw} and LHCb \cite{Aaij:2013noa}.
Throughout this paper, take the $b$-quark pole mass to be $m_b = 4.5$~GeV,
evaluate $\alpha_s^{(n_f)}(\mu_R)$ at NLO
with $n_f = 4$ and $\Lambda_{\overline{\mathrm{MS}}}^{(4)} = 328$~MeV
if $\mu_R<m_b$ and
with $n_f = 5$ and $\Lambda_{\overline{\mathrm{MS}}}^{(5)} = 226$~MeV
if $\mu_R>m_b$, and use the
CTEQ6.6M proton PDFs \cite{CTEQ6.6} unless otherwise stated.

\begin{figure}[t!] 
\begin{center}
\includegraphics[scale=0.63]{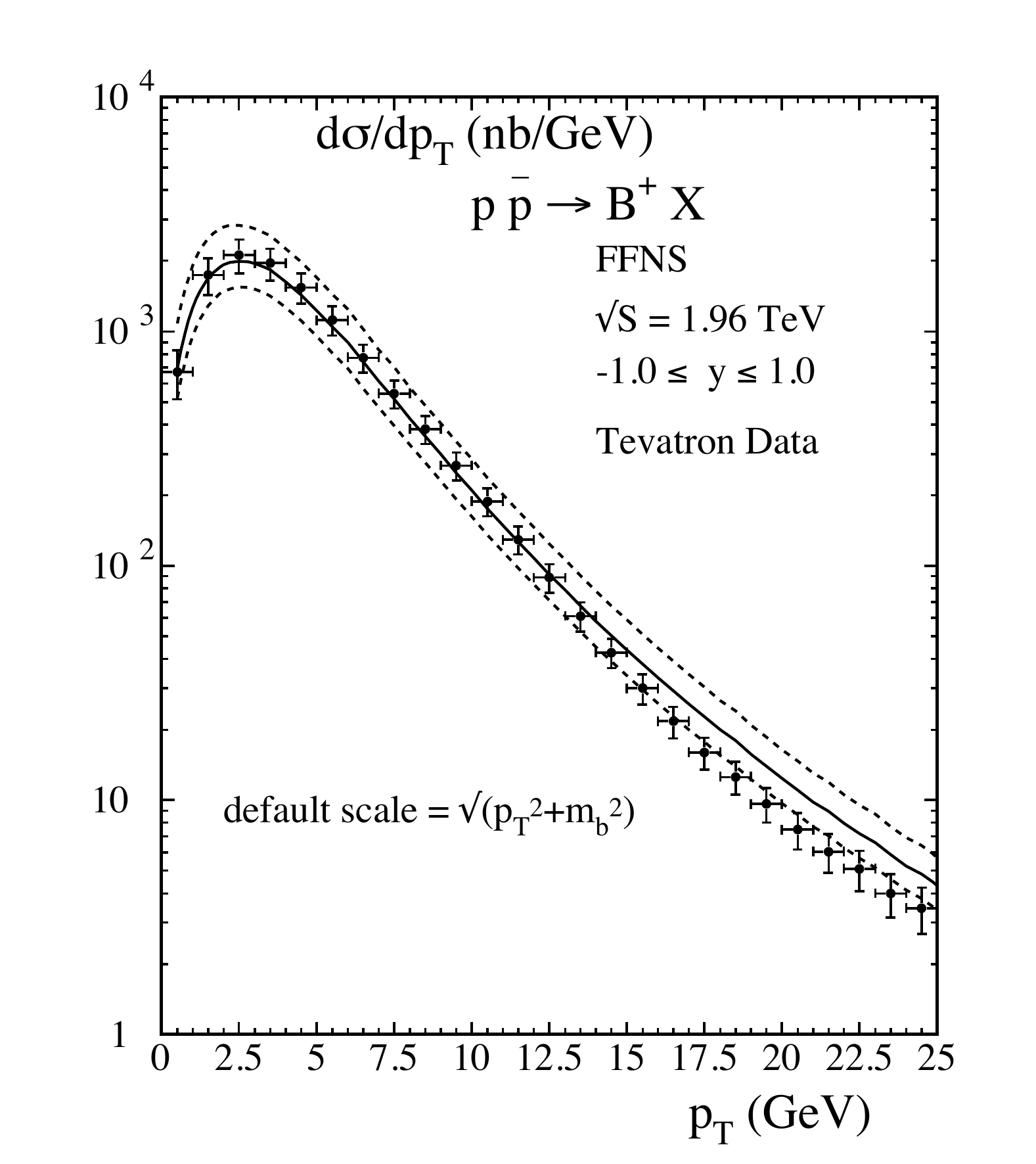}
\end{center}
\caption{
NLO predictions for $d\sigma/dp_T$ of $p\bar{p}\to B^++X$ with
$\sqrt{S}=1.96$~TeV and $-1\le y\le 1$ in the FFNS
are compared with CDF data \cite{Acosta:2004yw}.
The dashed lines represent the theoretical uncertainty estimated by varying 
$\mu_R$ and $\mu_I$ up and down by a factor of 2 
about the default choice $\mu_R = \mu_I = m_T$.
}
\label{fig1}
\end{figure}

We start with results to be compared with the CDF data \cite{Acosta:2004yw}.
In Fig.~\ref{fig1},
we show NLO predictions in the FFNS with $n_f = 4$. The full 
line shows the result for the default choice of scales,
$\mu_i = \xi_i m_T$ with $\xi_i = 1$ for $i=R,I$,
while the dashed lines represent an 
estimate of the theoretical error obtained in the usual way, by
varying $\xi_i$ up and down by a factor of 2. We take the transition 
of $b$ and $\bar{b}$ quarks to the observed $B$ mesons 
into account by using the branching fraction $B(b \to B) = 39.8\%$ 
\cite{Nakamura:2010zzi} as an overall normalization factor. The 
prediction in the FFNS agrees with the CDF data quite well, within 
experimental errors, up to $p_T \simeq 15$~GeV. Beyond this value 
of $p_T$, the FFNS starts to overestimate the data, as has been 
shown already in our previous publication \cite{Kniehl:2008zza}. 

\begin{figure}[t!] 
\begin{center}
\includegraphics[scale=0.43]{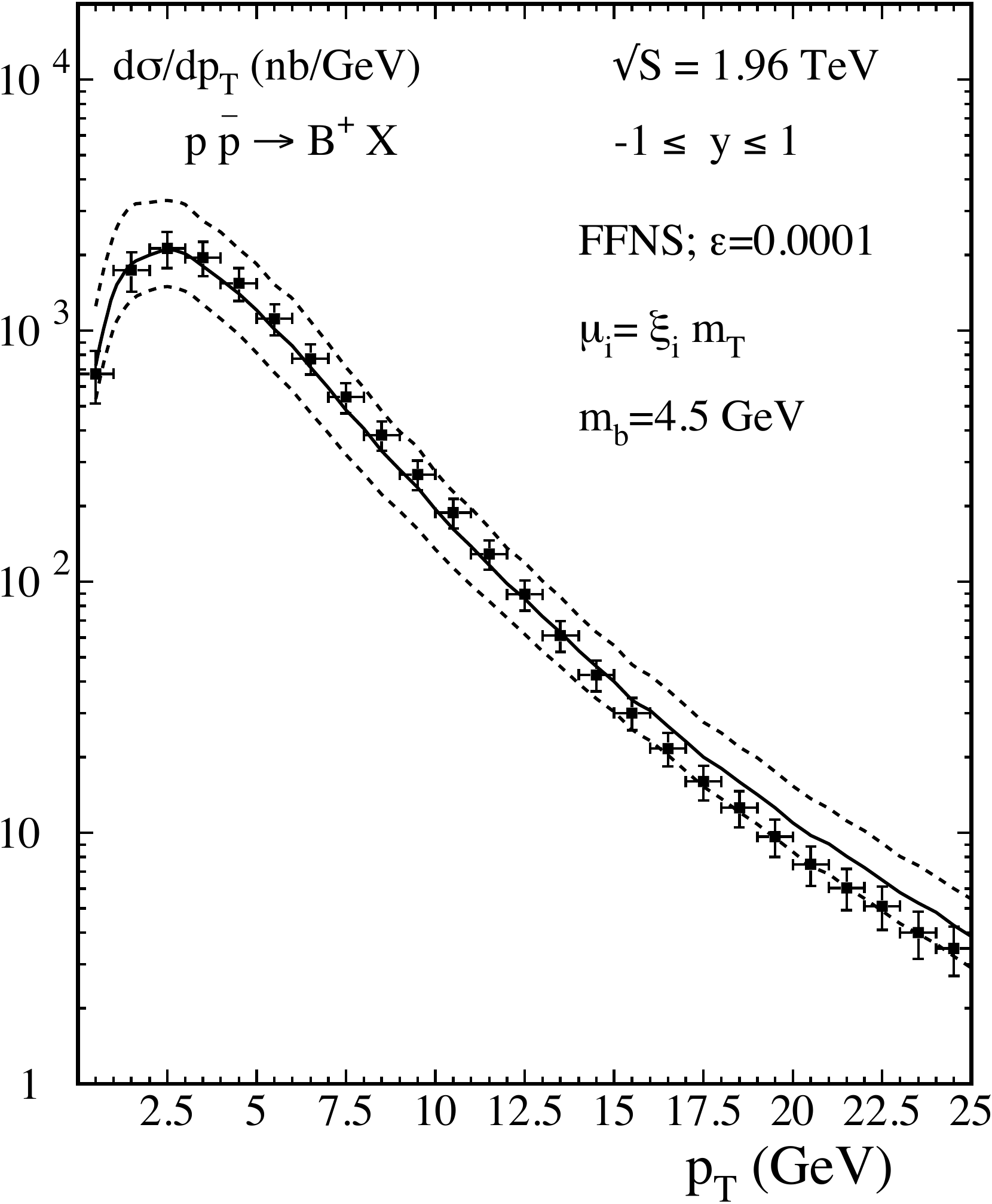}
\includegraphics[scale=0.43]{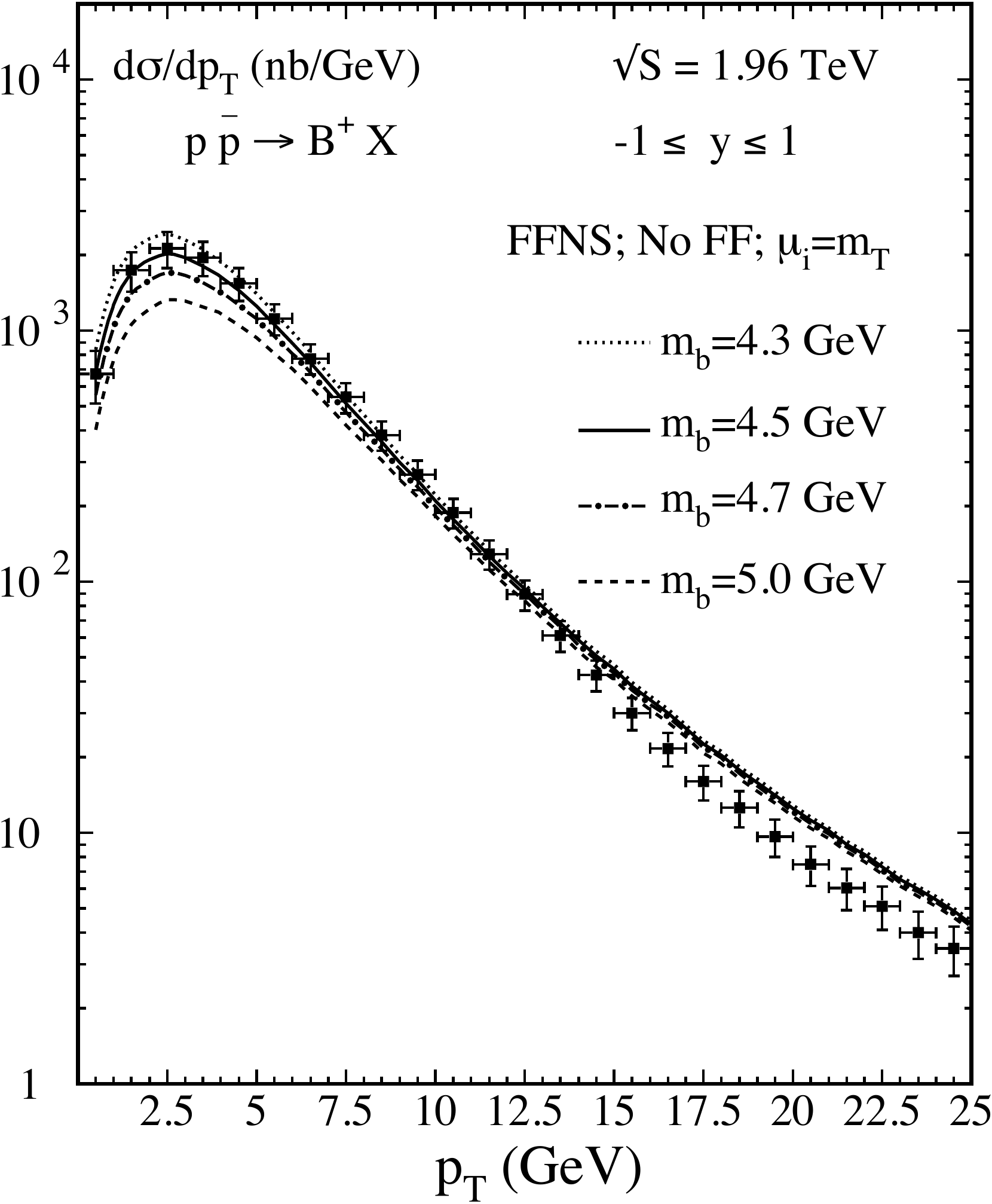}
\end{center}
\caption{
NLO predictions for $d\sigma/dp_T$ of $p\bar{p}\to B^++X$ with
$\sqrt{S}=1.96$~TeV and $-1\le y\le 1$ in the FFNS
are compared with CDF data \cite{Acosta:2004yw}.
Left panel: evaluation using the Peterson FF with $\epsilon = 10^{-4}$
and varying $\mu_R$ and $\mu_I$ by a factor 
of 2 up and down about the default choice $\mu_R = \mu_I 
= m_T$. Right panel: uncertainties from variations of $m_b$. 
}
\label{fig1a}
\end{figure}

In the FFNS, there is no need for FFs. However, a $\mu_F$-independent FF 
might be introduced on phenomenological 
grounds and because of theoretical considerations to guarantee a 
proper matching between the schemes with $n_f=4$ and $n_f=5$. 
In the left panel of Fig.~\ref{fig1a}, we show results obtained using the
$\mu_F$-independent Peterson FF \cite{Peterson:1982ak} with parameter
$\epsilon = 10^{-4}$. We find only marginal differences 
with respect to the case where a constant branching fraction is used. Note 
that there are no $g,q,\bar{q} \to B$ transitions in the 
FFNS. 

In addition to uncertainties from scale variations, there are 
also uncertainties due to errors in the input. 
We postpone the discussion of errors in the parametrizations of the
PDFs to when we present predictions for the LHCb experiments, in 
Fig.~\ref{fig6} below, but instead show the influence of $m_b$
variations on the default prediction for the Tevatron measurements in the
right panel of Fig.~\ref{fig1a}. At small $p_T$ values, the uncertainty 
is comparable in size with the scale uncertainty, but it is 
negligible for $p_T\agt2m_b$.

\begin{figure}[t!] 
\begin{center}
\raisebox{3.3mm}{
\includegraphics[scale=0.63]{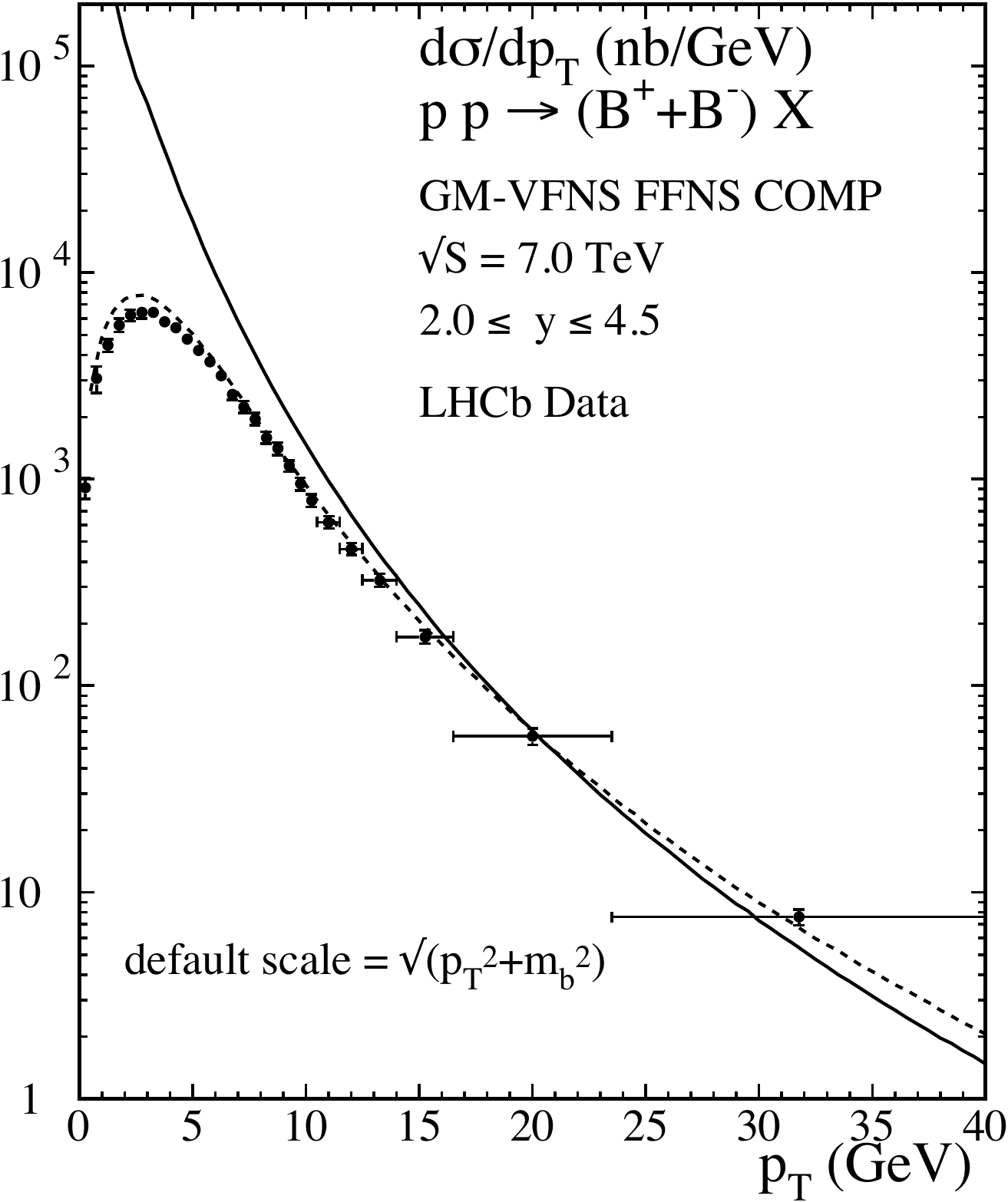}
}
\end{center}
\caption{
NLO predictions for $d\sigma/dp_T$ of $pp\to B^\pm+X$ with
$\sqrt{S}=7$~TeV and $2\le y\le4.5$ in the FFNS
(dashed line) and the GM-VFNS (solid line) are compared with LHCb data
\cite{Aaij:2013noa}.
}
\label{fig2}
\end{figure}

In Ref.~\cite{Kniehl:2008zza}, we presented detailed 
comparisons of FFNS and GM-VFNS results with CDF 
data for $B^+$ production (see, e.g., Figs.~7 and 8 in
Ref.~\cite{Kniehl:2008zza}). For calculations in the GM-VFNS, we use 
the $\mu_F$-dependent FFs described in Ref.~\cite{Kniehl:2008zza}.
In this case, there are also small contributions due to transitions from light
quarks and gluons to $B$ mesons. With the default choice
$\xi_R = \xi_I = \xi_F= 1$, the GM-VFNS predictions 
diverge for $p_T \to 0$, in obvious disagreement with the data. 
We notice, however, that the FFNS and GM-VFNS predictions approach each other 
at around $p_T \simeq 20$~GeV, i.e.\ 4 to 5 times $m_b$. In Fig.~\ref{fig2},
we show a similar comparison of the FFNS and GM-VFNS 
predictions for the case of the LHCb data \cite{Aaij:2013noa}. 
Although these data correspond to much higher $\sqrt{S}$ values and to
different $y$ ranges compared with the 
previous results in Ref.~\cite{Kniehl:2008zza}, we observe similar 
qualitative behaviors of the FFNS and GM-VFNS predictions 
and a transition point at about the same value of $p_T$, namely 
$p_T\simeq 20$~GeV.

\begin{figure}[t!] 
\begin{center}
\includegraphics[scale=0.63]{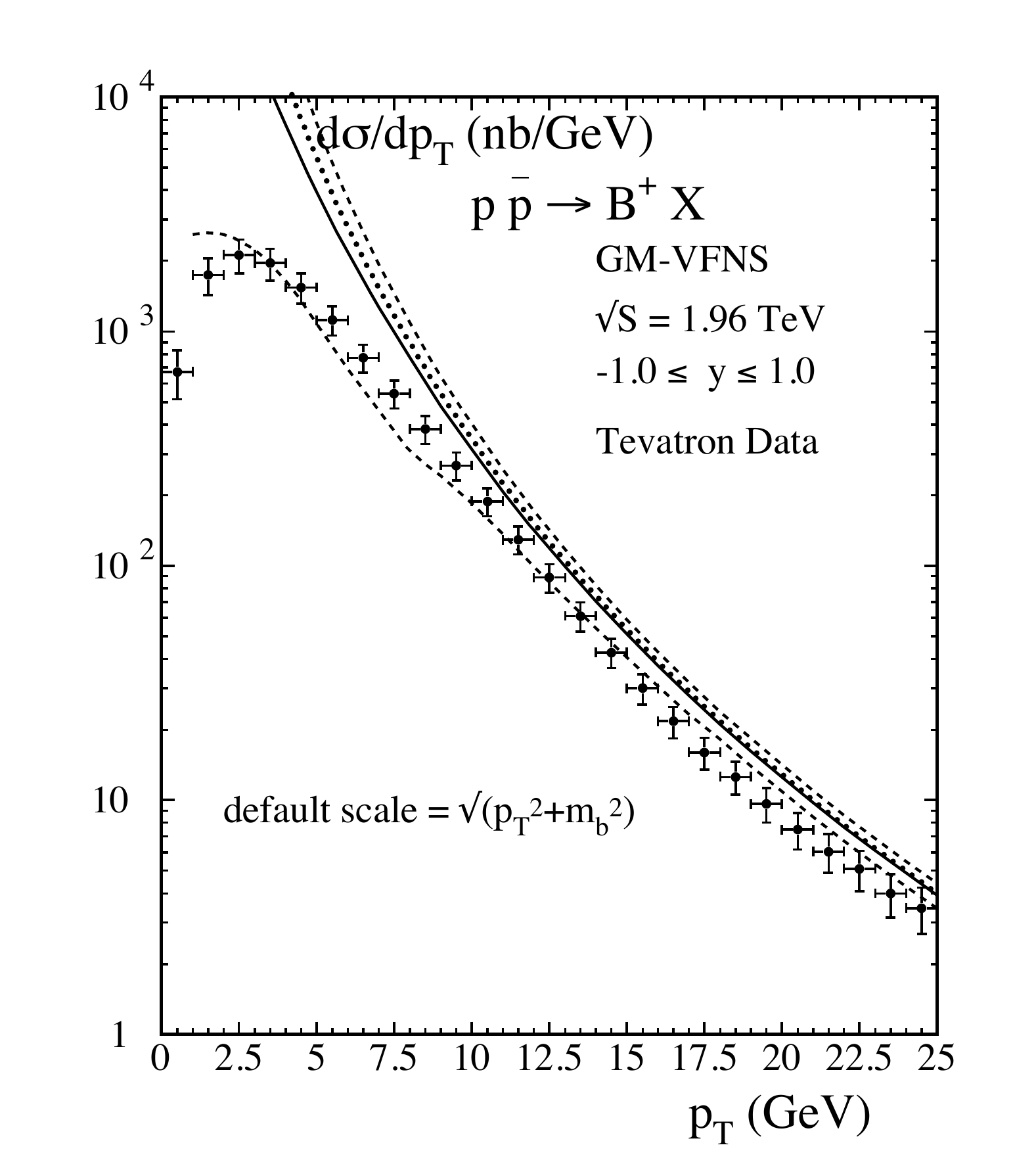}
\end{center}
\caption{
NLO predictions for $d\sigma/dp_T$ of $p\bar{p}\to B^++X$ with
$\sqrt{S}=1.96$~TeV and $-1\le y\le 1$ in the GM-VFNS
are compared with CDF data \cite{Acosta:2004yw}.
Besides the default prediction for scale parameters $\xi_i = 1$ (solid line),
a theoretical-error band encompassed by the predictions for $\xi_i = 1/2$
(upper dashed line) and $\xi_i = 2$ (lower dashed line) is shown.
The effect of including quark and hadron mass corrections via the phase space
in the default prediction is also shown, for $M_H = 5.28$~GeV (dotted line).
}
\label{fig3}
\end{figure}

One should expect that a correct treatment of kinematic 
constraints due to the finite heavy-hadron mass is important, 
in particular at small $p_T$ values. However, there is no unique 
prescription to take into account the finite masses of the heavy quark and
the heavy hadron at the same 
time. A prescription for including the heavy-hadron mass in connection with
massless quarks based on the light-cone scaling variable was introduced in
Ref.~\cite{Albino:2008fy}.
Here, we propose a generalization to the massive-quark case.
This amounts to scaling the partonic cross section for the production of the
massive quark $a$, with mass $m_a$, energy $E_a$, and three-momentum $p_a$,
that initiates the formation of the heavy hadron $H$, with mass $M_H$, energy
$E_H=\sqrt{M_H^2+p_T^2}\cosh y$, and three-momentum
$p_H=\sqrt{M_H^2\sinh^2y+p_T^2\cosh^2y}$, as
\begin{equation}
{\rm d}\sigma_a \rightarrow \frac{{\rm d}\sigma_a}{R_{a\to H}^2},
\qquad
R_{a\to H} = 1 - \frac{M_H^2 - z^2 m_a^2}{(E_H+p_H)^2 - z^2 m_a^2} \, ,
\end{equation}
where $z=(E_H + p_H)/(E_h + p_h)$ is the light-cone scaling variable.
In contrast to naive expectations, one finds $R_H < 1$ corresponding to a
slight enhancement of the cross section. In Fig.~\ref{fig3}, we show 
the result of a calculation where this correction factor is taken 
into account. The effect is small mainly because it is only the 
difference between the $b$-quark and $B$-meson masses 
that enters. Therefore, also the additional suppression from 
tighter phase space limits is numerically not relevant. The poor
small-$p_T$ behavior of the GM-VFNS calculation cannot be remedied 
by such a naive treatment of phase space restrictions. 
%
%

\begin{figure}[t!] 
\begin{center}
\includegraphics[scale=0.5]{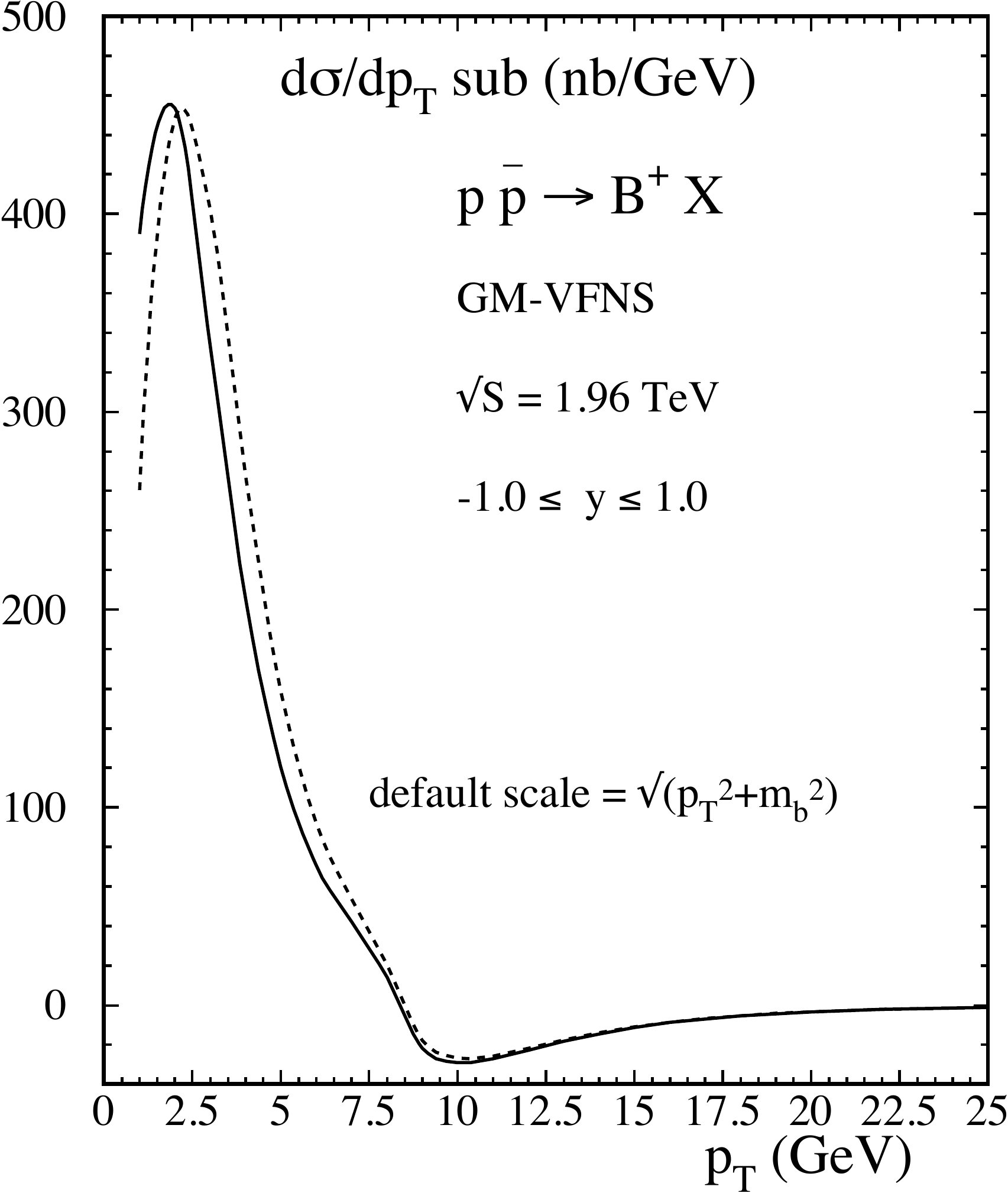}
\end{center}
\caption{
Subtraction term for the $gg$ channel evaluated with zero (solid line) and
finite (dashed line) $m_b$ value in the NLO prediction for
$d\sigma/dp_T$ of $p\bar{p}\to B^++X$ with $\sqrt{S}=1.96$~TeV and
$-1\le y\le 1$ in the GM-VFNS.
}
\label{fig5a}
\end{figure}

The results shown in Figs.~\ref{fig1}--\ref{fig3} suggest that 
the predictions of the FFNS and the GM-VFNS are very similar in 
the range 15~GeV${}\alt p_T \alt 20$~GeV. In this range, both 
theories are justified , with the FFNS naturally extending to 
smaller $p_T$ values and the GM-VFNS to larger $p_T$ values.
Therefore, one could designate a transition 
point $\mu_t$ at $p_T \simeq 20$~GeV, say, where one switches from one 
scheme to the other \cite{Collins:1998rz,Tung:2001mv}. For 
practical purposes, this would require the knowledge of the 
PDFs in the scheme with $n_f = 4$ active flavors 
up to $\mu_I = \mu_t$ \cite{Kusina:2013slm}. A naive prescription 
to combine the two schemes could be to use matching functions 
$\theta(\mu_t^2 - p_T^2)$ and $[1- \theta(\mu_t^2 - p_T^2)]$ to multiply the
FFNS and GM-VFNS cross sections, respectively.
Such a prescription will, however, lead 
to a discontinuity in the combined cross section 
at $p_T=\mu_t$, a property that is certainly unphysical. 
One could try to invent different ways to combine the two 
schemes by introducing some smooth matching function, e.g.,
$G(p_T) = p_T^2 / (p_T^2 + c^2m_b^2)$ with $c=5$ \cite{Cacciari:1998it}, and 
imposing, schematically, the prescription $\sigma = G(p_T) 
\sigma_{\mathrm{GM-VFNS}} + [1-G(p_T)] \sigma_{\mathrm{FFNS}}$.
However, such an approach would introduce a new parameter, and it is 
unclear how theoretical uncertainties related to the choice of 
this parameter should be estimated.

It will be helpful to take a closer look at the structure of the 
partonic cross sections in the GM-VFNS. They involve 
subtraction terms needed to separate the large logarithms 
$\ln (p_T^2 / m_b^2)$ at fixed relative order $\alpha_s$. For 
example, the cross section of the partonic subprocess $gg \rightarrow 
gb\bar{b}$ contains terms with large logarithms which can be written, 
schematically, as convolutions $f(g \rightarrow b) \otimes 
\sigma(gb \rightarrow gb)$ and $\sigma(gg \rightarrow gg)
\otimes d(g \rightarrow b)$ with a partonic PDF 
$f(g \rightarrow b)$ and a partonic FF $d(g \rightarrow b)$ 
(see Ref.~\cite{Kniehl:2005mk} for definitions and a precise 
formulation of the corresponding cross sections). These splitting 
functions are absorbed in the PDFs and FFs, respectively. The large 
logarithms subtracted in this way are added back by calculating the 
cross sections of the $gb \rightarrow gb$ and  $gg \rightarrow gg$
subprocesses with a $b$-quark PDF and a $g \rightarrow B$ FF, respectively.
There are similar subtraction terms and corresponding contributions 
involving PDFs and FFs of light quarks and the gluon in other 
channels as well. 

In Fig.~\ref{fig5a}, we show numerical results for the subtraction 
terms needed in the $gg$ channel (see Eqs.~(45)--(53) in
Ref.~\cite{Kniehl:2005mk}), using a linear scale for better visibility.
Their contribution is small, but not negligible, even at large $p_T$ values.
Specifically, the evaluations using the correct prescription with
$m_b \neq 0$ (dashed line) and the approximate one with $m_b = 0$ (full line)
are compared with each other.
We observe from this that such mass effects are small and cannot be responsible
for the unphysical increase of the total result for $d\sigma / dp_T$ towards
small $p_T$ values.

It is, of course, unavoidable that the subtracted terms differ 
from those added back in the PDFs and FFs. The subtractions 
are obtained at fixed order, $O(\alpha_s)$ in our case, 
while the PDFs and FFs contain the large logarithms resummed 
to all orders. Formally, the differences are of higher order 
in $\alpha_s$. However, these higher-order terms are folded with 
cross sections calculated in the ZM-VFNS, and these 
cross sections are singular for $p_T \rightarrow 0$. Therefore, it 
is not surprising that the contributions with $b$ quarks in the 
initial state dominate at small $p_T$ values. 

Obviously, NLO cross sections of $b$-quark-initiated subprocesses that are
convoluted with PDFs and FFs evolved at NLO contain terms singular
for $p_T \rightarrow 0$ at one order beyond the subtracted terms. In 
fact, it would be a major task to derive the missing
next-to-next-to-leading-order (NNLO) 
subtraction terms. With such an extended version of the GM-VFNS 
including NNLO subtractions, but still using partonic cross 
sections derived in the ZM-VFNS, the problem would be 
shifted to one order higher, but it remains to be seen whether 
the required cancellations of singular terms can be obtained with 
the required numerical precision. In addition, strictly speaking, 
the NNLO subtraction terms only make sense in combination with 
the fixed-order calculation at NNLO. Otherwise, at large $p_T$ values, 
the NNLO subtraction terms and their NLO fixed-order counterparts do not
cancel, and the ZM-VFNS is not recovered. 

In turn, one could argue that $b$-quark-initiated processes 
evaluated with $b$-quark PDFs and FFs should be included 
in the GM-VFNS at LO only. The predictions thus obtained \cite{Olness:1997yc}
exhibit better agreement with data in the medium-$p_T$ range, between 2 and
7~GeV or so, but switching off these NLO terms is again 
not sufficient to completely eliminate a singular behavior  
for $p_T \rightarrow 0$. Moreover, and most importantly, numerical 
evaluations show that NLO corrections in the zero-mass part of 
the GM-VFNS are essential at large $p_T$ values. Therefore, we do not 
follow this option either.

Instead, we try to exploit the freedom offered by the 
presence of $\mu_R$, $\mu_I$, and $\mu_F$, 
parameters that are present anyway. Their values are not determined 
by theory, but some choice has to be made, based on some reasonable, 
but ad-hoc physical argument. In fact, a judicious choice of scales 
can lead to a suppression of the potentially dangerous contributions 
from initial-state $b$ quarks. This exploits the fact 
that all commonly available PDF fits assume that the $b$-quark PDF 
is zero below some starting scale, usually chosen to be $\mu_I = m_b$.
The same is true for the FFs: the 
FF for the $b \to B$ transition vanishes for $\mu_F < m_b$. 
Therefore, with $\mu_{I,F} = \xi_{I,F} m_T$, a value $\xi_{I,F}<1$ will
render the $b$-quark PDF and FF zero for $p_T < m_b \sqrt{1/\xi_{I,F}^2-1}$. 

\begin{figure}[t!] 
\begin{center}
\includegraphics[scale=0.45]{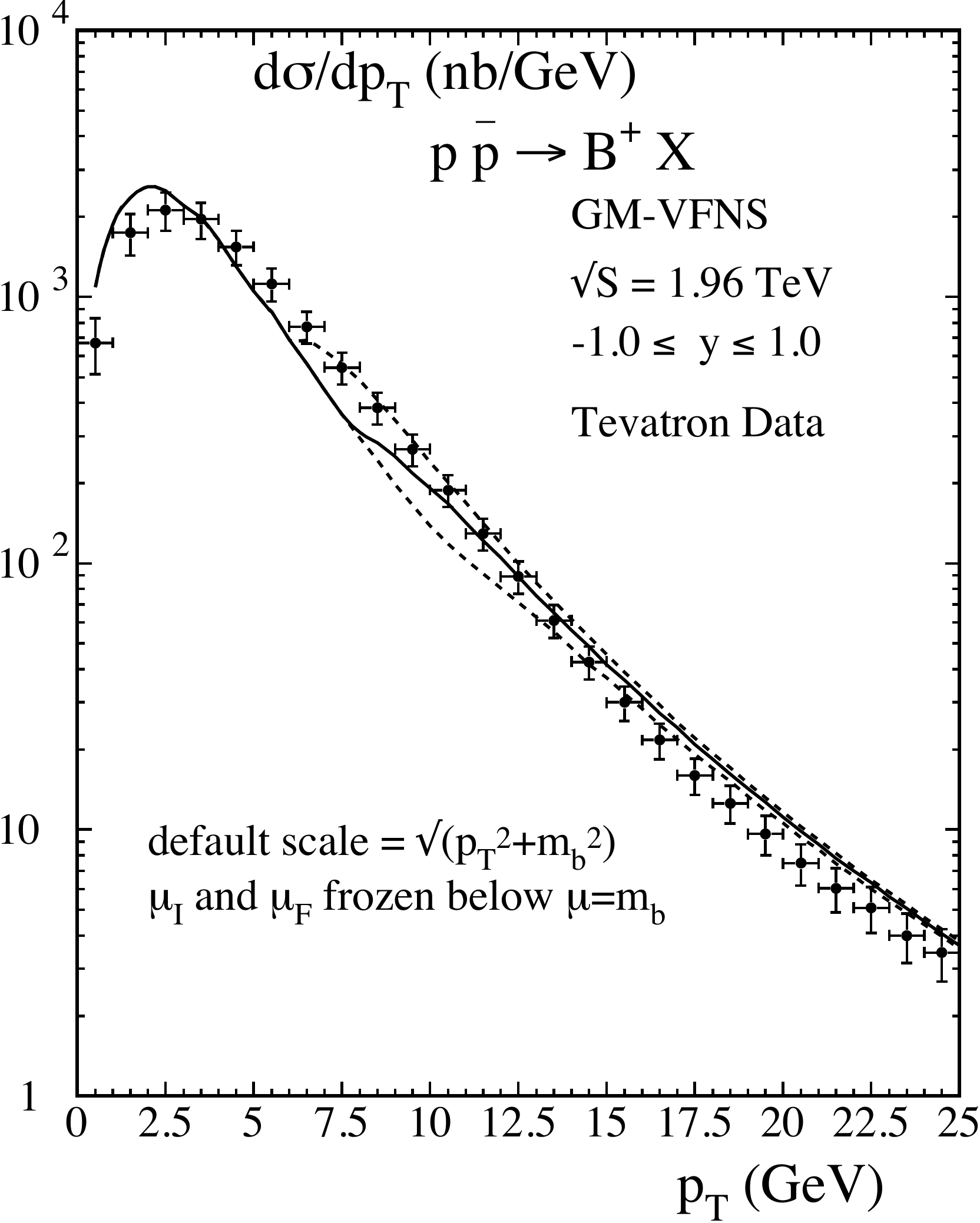}
\includegraphics[scale=0.45]{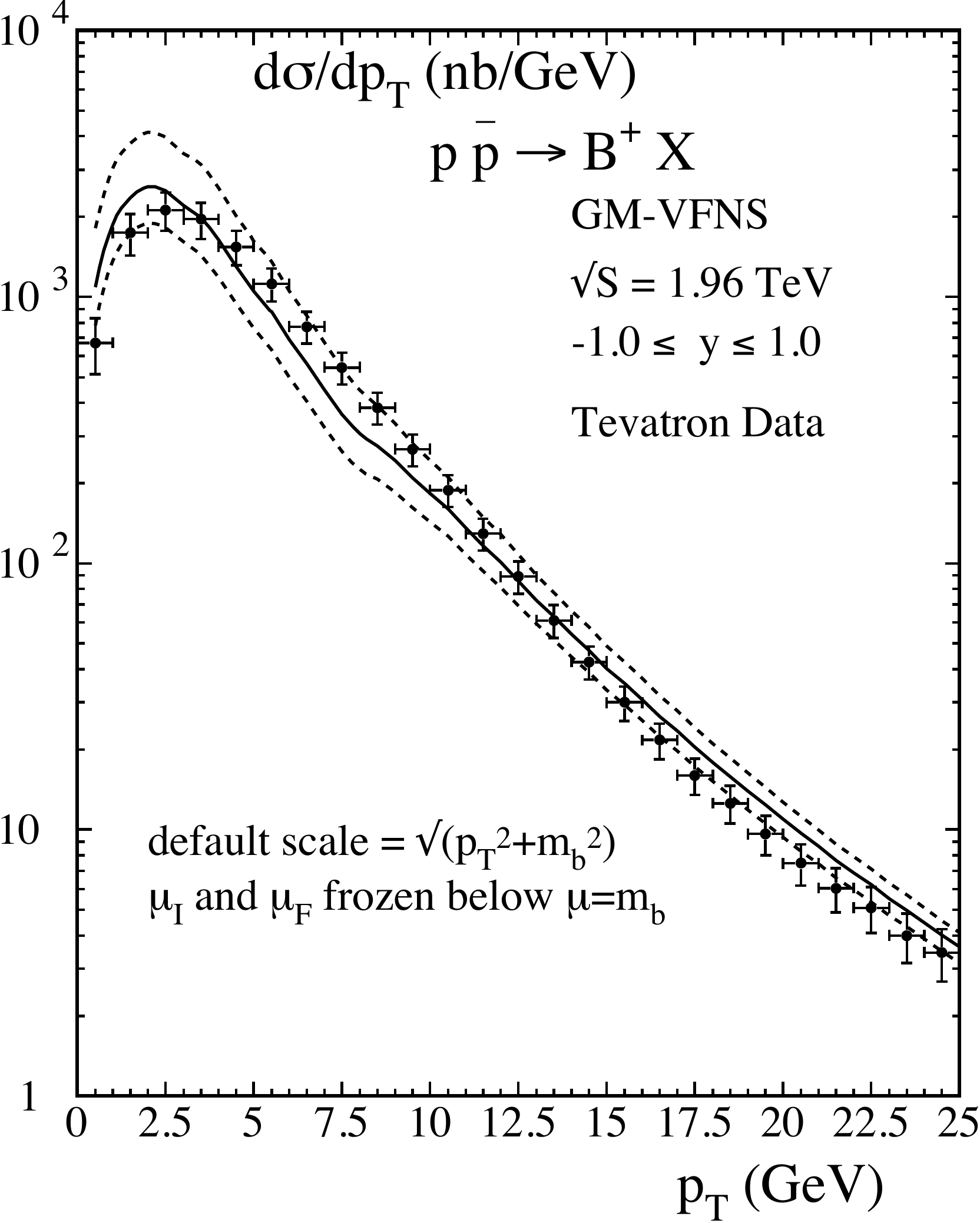}
\end{center}
\caption{
NLO predictions for $d\sigma/dp_T$ of $p\bar{p}\to B^++X$ with
$\sqrt{S}=1.96$~TeV and $-1\le y\le 1$ in the GM-VFNS are
compared with CDF data \cite{Acosta:2004yw}.
Left panel: evaluations with $\xi_R = 1$ and  $\xi_I=\xi_F = 0.5$,
(solid line), 0.4 (lower dashed line), and 0.6 (upper dashed line). 
Right panel: evaluations with $\xi_I=\xi_F = 0.5$ and 
$\xi_R = 1$ (solid line), 0.5 (upper dashed line), and 2 (lower dashed line).
If $\xi_{I,F}<1$, then $\mu_{I,F}=m_b$ is put for
$p_T<m_b\sqrt{1/\xi_{I,R}^2-1}$. 
}
\label{fig4ab}
\end{figure}

In Fig.~\ref{fig4ab}, we show results obtained with such scale choices and
compare them with CDF data \cite{Acosta:2004yw}.
Indeed, values for $\xi_I$ and $\xi_F$ of about 1/2 lead to the required
suppression of $b$-quark-initiated contributions.
Specifically, in the left panel of Fig.~\ref{fig4ab}, we choose
$\xi_R = 1$ and $\xi_I = \xi_F = 0.4, 0.5, 0.6$.
In all these cases, there is a turn-over near $p_T = 2.5$~GeV, and the
agreement with the CDF data is reasonably good.
We impose the additional constraint 
that $\mu_I$ and $\mu_F$ are not allowed to take values $\mu_I,\mu_F < m_b$,
i.e.\ the 
DGLAP evolutions of the PDFs and FFs are frozen below this 
scale.\footnote{In fact, we have 
  to freeze $\mu_I$ slightly below $m_b$, 
  $\mu_I = C m_b$ with $C<1$, since the $b$-quark PDF parametrization of the
  CTEQ6.6M set, which we use here, vanishes only strictly below threshold.
For our numerical 
  evaluations, we choose $C=0.99$.} This explains the slight 
bumps that occur in the $p_T$ distribution at $p_T = 6.0$, 7.8, and 
10.3~GeV for $\xi_I=\xi_F = 0.6$, 0.5, and 0.4, respectively. 
Obviously, the freedom in the choice of the default values of 
$\xi_I$ and $\xi_F$ leads to a moderate extra uncertainty close to the 
transition region, but the results in the small-$p_T$ range,
$p_T \alt 5$~GeV, and for large $p_T$ values stay unaffected. 

Taking now $\xi_R = 1$, $\xi_I = \xi_F = 0.5$ as the new default 
scale choice, we estimate the theoretical errors in the 
usual way by varying the scale parameters by a factor 2 up and 
down about the default scale choice. In fact, it turns out that the 
variation of $\mu_R$ is the dominant source 
of the theoretical uncertainties, and we simplify the subsequent 
calculations by only considering variations of $\xi_R$ in the 
range from 0.5 to 2. Note that we do not introduce an extra 
prescription to freeze $\mu_R$ below $m_b$ because, first, the choice of
$\mu_R$ is not related to switching off $b$-quark-initiated 
subprocesses and, second, full variations of $\mu_R$ are 
needed to obtain realistic estimates of the theoretical 
uncertainty. The resulting error band is shown 
in the right panel of Fig.~\ref{fig4ab}. We emphasize that the freedom 
in choosing specific values of $\xi_I$ and $\xi_F$ as default does not 
introduce a large additional uncertainty, as may be understood by 
comparing the left and right panels of Fig.~\ref{fig4ab}. 

\begin{figure}[h!] 
\begin{center}
\includegraphics[scale=0.57]{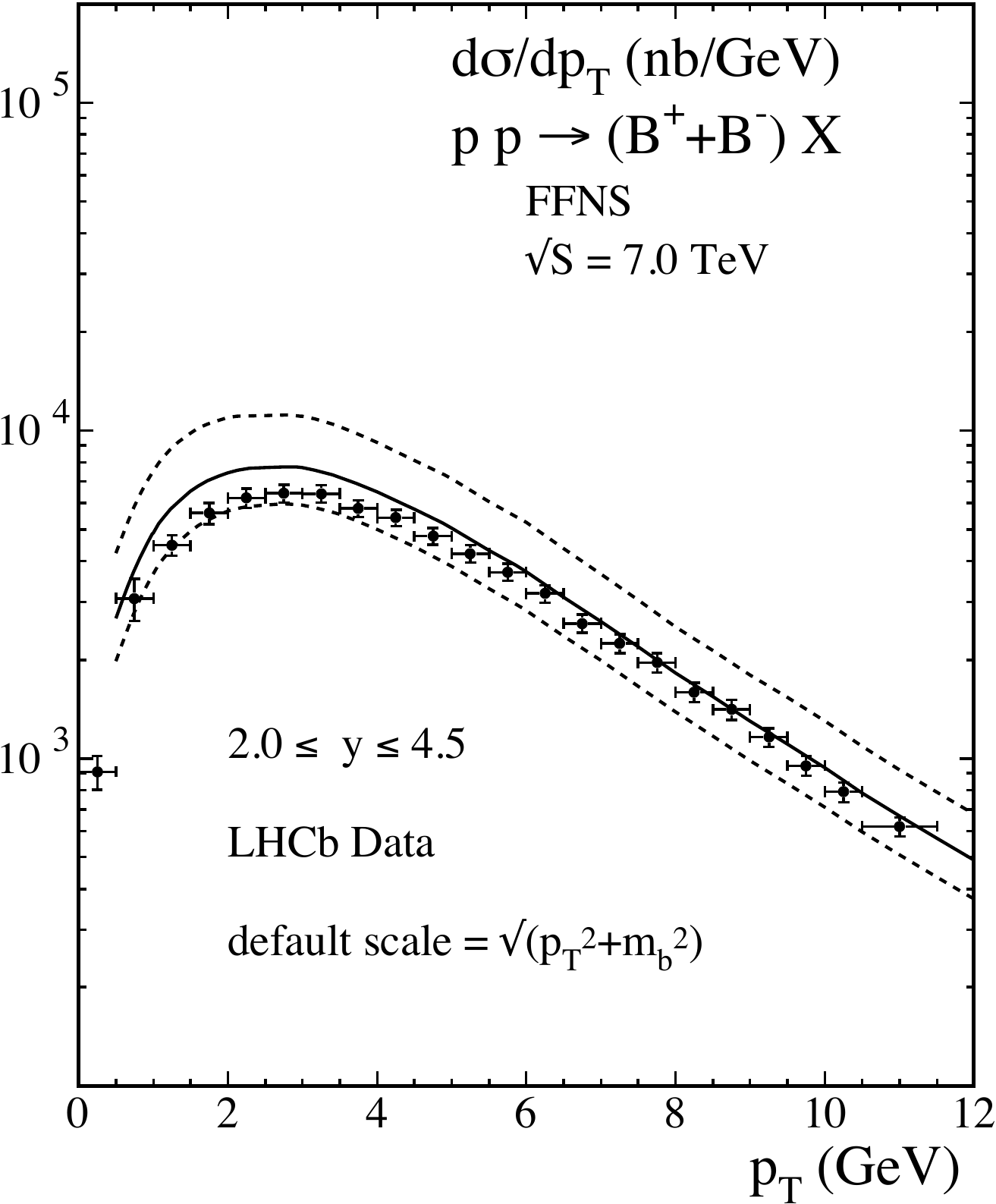}
\includegraphics[scale=0.45]{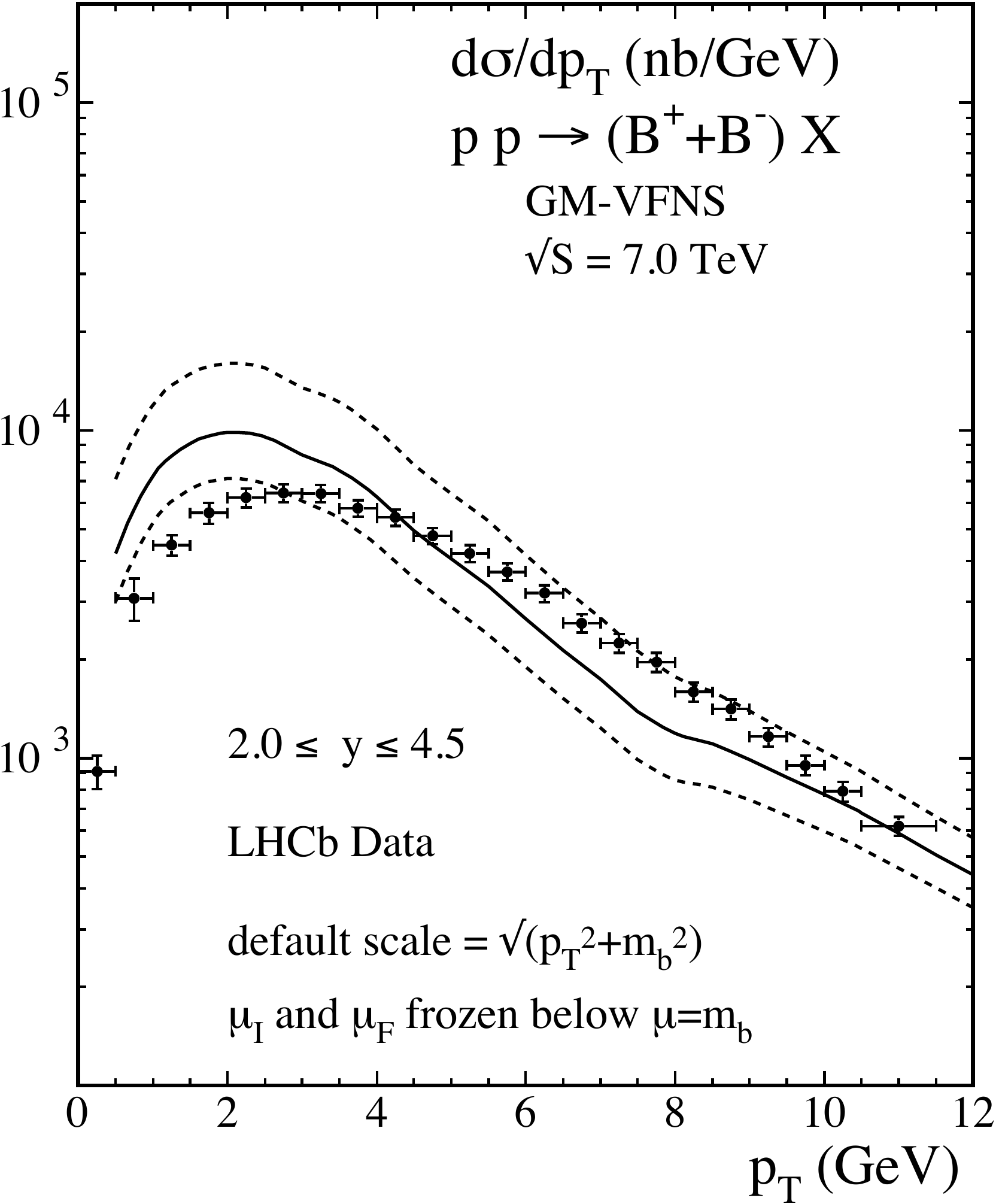}
\end{center}
\caption{
NLO predictions for $d\sigma/dp_T$ of $pp\to B^\pm+X$ with
$\sqrt{S}=7$~TeV and $2\le y\le4.5$ in the FFNS (left panel)
and the GM-VFNS (right panel) are compared with LHCb data
\cite{Aaij:2013noa}.
The default scale choice is $\xi_R=1$ and $\xi_I=\xi_F=0.5$ (solid lines), and
the theoretical-error bands are obtained by varying $\xi_R$ by a factor of 2
up (lower dashed lines) and down (upper dashed lines).
If $\xi_{I,F}<1$, then $\mu_{I,F}=m_b$ is put for
$p_T<m_b\sqrt{1/\xi_{I,R}^2-1}$. 
}
\label{fig5ab}
\end{figure}

In Fig.~\ref{fig5ab}, we compare the LHCb data \cite{Aaij:2013noa} with our
theoretical predictions implemented with the scale conventions introduced
above.
In the 
left panel, we see again nice agreement between the FFNS predictions and 
experimental data. The right panel tells us that the data are not quite as 
well described by the GM-VFNS predictions, but the agreement 
is quite acceptable, as the data points are covered by the error band 
for $p_T\agt 2.5$~GeV. In this $p_T$ range, 
the agreement of the data with predictions obtained in the FONLL scheme 
\cite{Cacciari:2012ny} is quite similar \cite{Aaij:2013noa}.

In Fig.~\ref{fig6}, we show a comparison of results obtained using 
different PDF parametrizations.
Specifically, we use the CT10 \cite{Lai:2010vv},
MSTW \cite{Martin:2009bu},
HERA-PDF1.5(NLO) \cite{HERAFitter}, and
NNPDF 2.3 \cite{Ball:2011mu,Ball:2011uy} PDF sets.
The CTEQ6.6M \cite{CTEQ6.6} PDFs adopted elsewhere in this paper yield a
result that is very similar to the one obtained using the CT10 PDFs.
We observe that there 
are differences in the small-$p_T$ range, $p_T\alt4$~GeV, 
which are somewhat larger than the experimental errors. We 
should, therefore, expect that $B$-meson production data 
at the LHC will help us to further constrain the PDFs. In particular, 
there is sensitivity to the gluon PDF in this kinematic 
range. 

\begin{figure}[h!] 
\begin{center}
\includegraphics[scale=0.49]{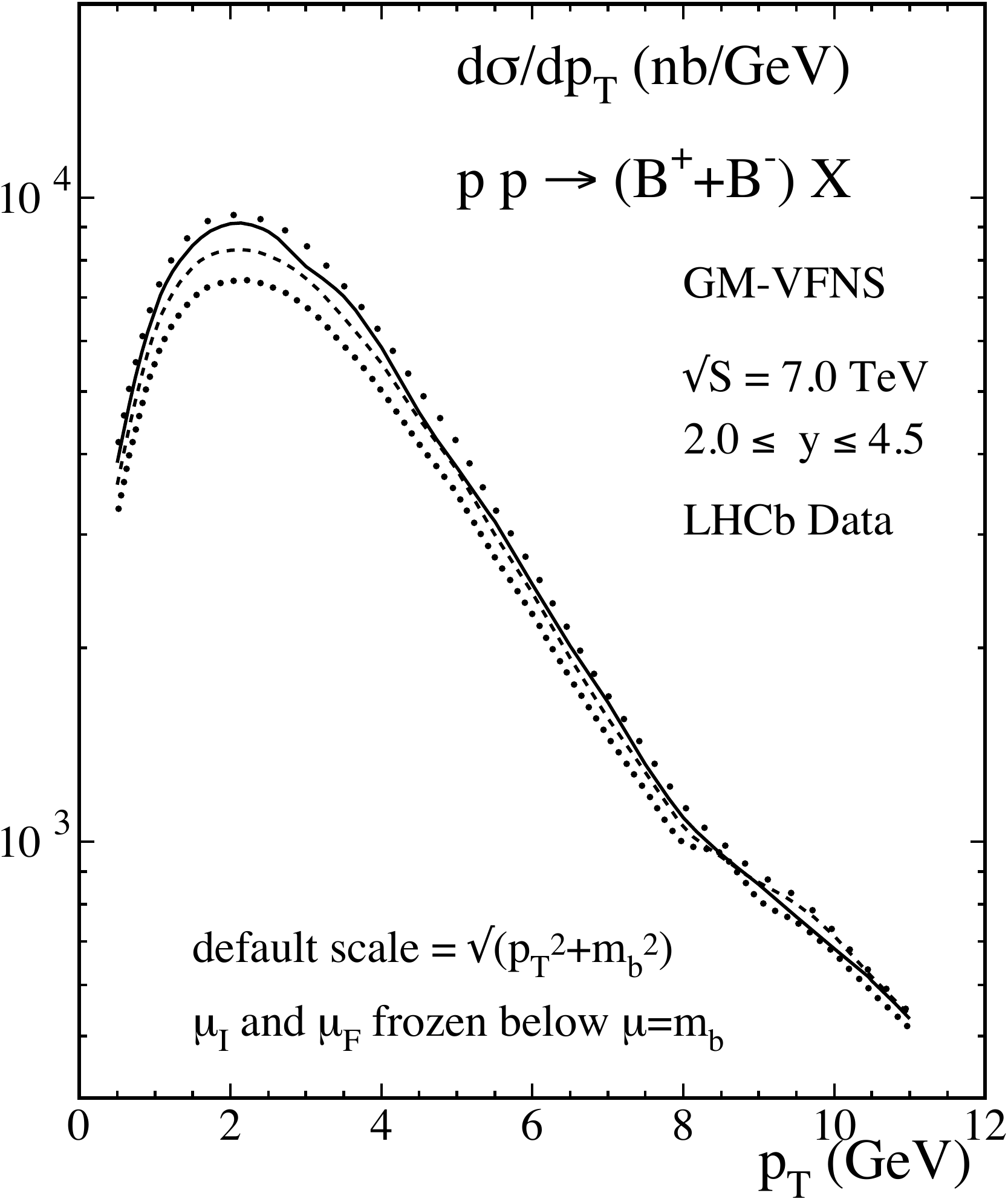}
\end{center}
\caption{
NLO predictions for $d\sigma/dp_T$ of $pp\to B^\pm+X$ with
$\sqrt{S}=7$~TeV and $2\le y\le4.5$ in the GM-VFNS with default scale choice
are compared with LHCb data \cite{Aaij:2013noa}.
The PDF error is estimated by using the CT10 (solid line) \cite{Lai:2010vv},
MSTW (dashed line) \cite{Martin:2009bu},
HERA\-PDF1.5(NLO) (lower dotted line) \cite{HERAFitter}, and
NNPDF 2.3 (upper dotted line) \cite{Ball:2011mu,Ball:2011uy} sets.
}
\label{fig6}
\end{figure}

\begin{figure}[h!] 
\begin{center}
\includegraphics[scale=0.46]{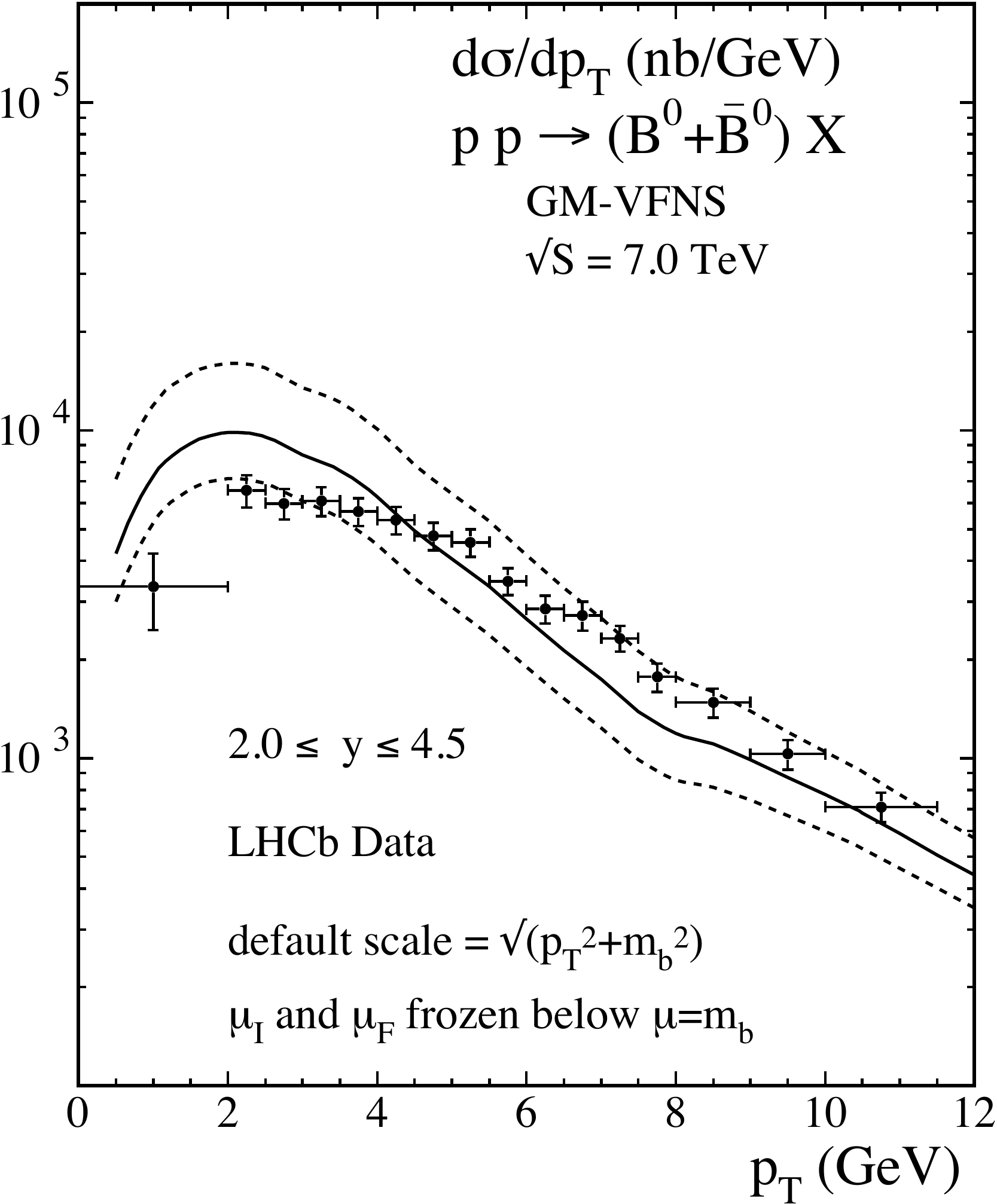}
\includegraphics[scale=0.46]{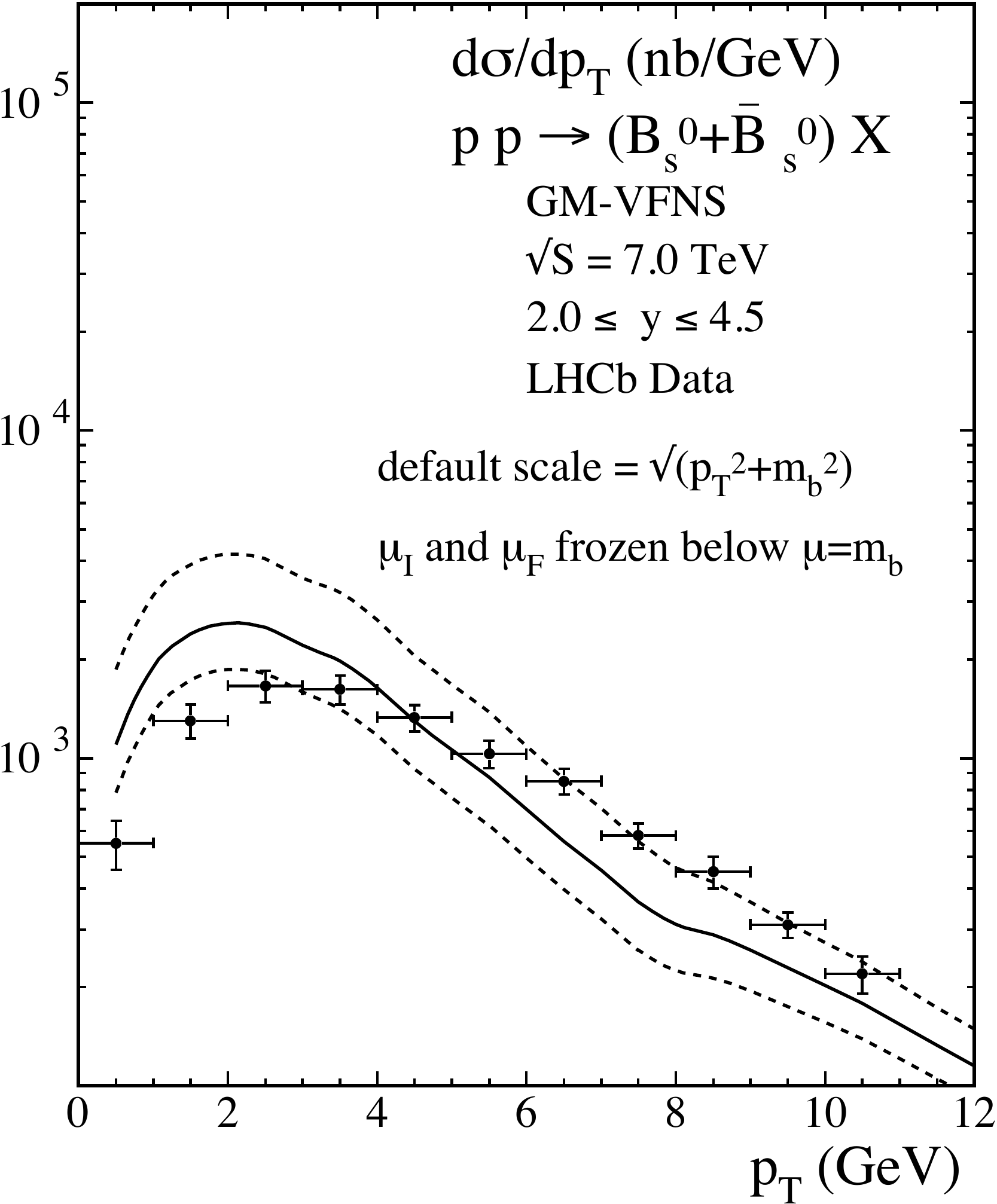}
\end{center}
\caption{
NLO predictions for $d\sigma/dp_T$ of
$pp\to B^0/\overline{B}^0+X$ (left panel) and
$pp\to B_s^0/\overline{B}_s^0+X$ (right panel)
with $\sqrt{S}=7$~TeV and $2\le y\le4.5$ in the GM-VFNS 
are compared with LHCb data \cite{Aaij:2013noa}.
The default predictions and theoretical-error bands are evaluated as
in the right panel of Fig.~\ref{fig4ab}. 
}
\label{fig7ab}
\end{figure}

For completeness, we also consider the production 
of $B^0$ and $B_s^0$ mesons. Appropriate experimental data were
published by the LHCb Collaboration in Ref.~\cite{Aaij:2013noa}. 
In Fig.~\ref{fig7ab}, we present comparisons with NLO predictions 
in the GM-VFNS using the scale setting and theoretical-error estimation
prescriptions described above to find good agreement.

Finally, we compare $B^+$-meson production data taken by the ATLAS
Collaboration \cite{ATLAS:2013cia} with our NLO GM-VFNS predictions.
These data extend into the very-large-$p_T$ range, where we expect the GM-VFNS 
to be quite appropriate.
Indeed, we find good agreement, except for the lowest $p_T$
bin, 9--13~GeV, and for central rapidities, where the data are slightly 
overestimated. Because of the large $p_T$ values probed, the 
adjustment of scales to match to the FFNS as described above 
is not an issue here. 

\begin{figure}[h!] 
\begin{center}
\includegraphics[scale=0.55]{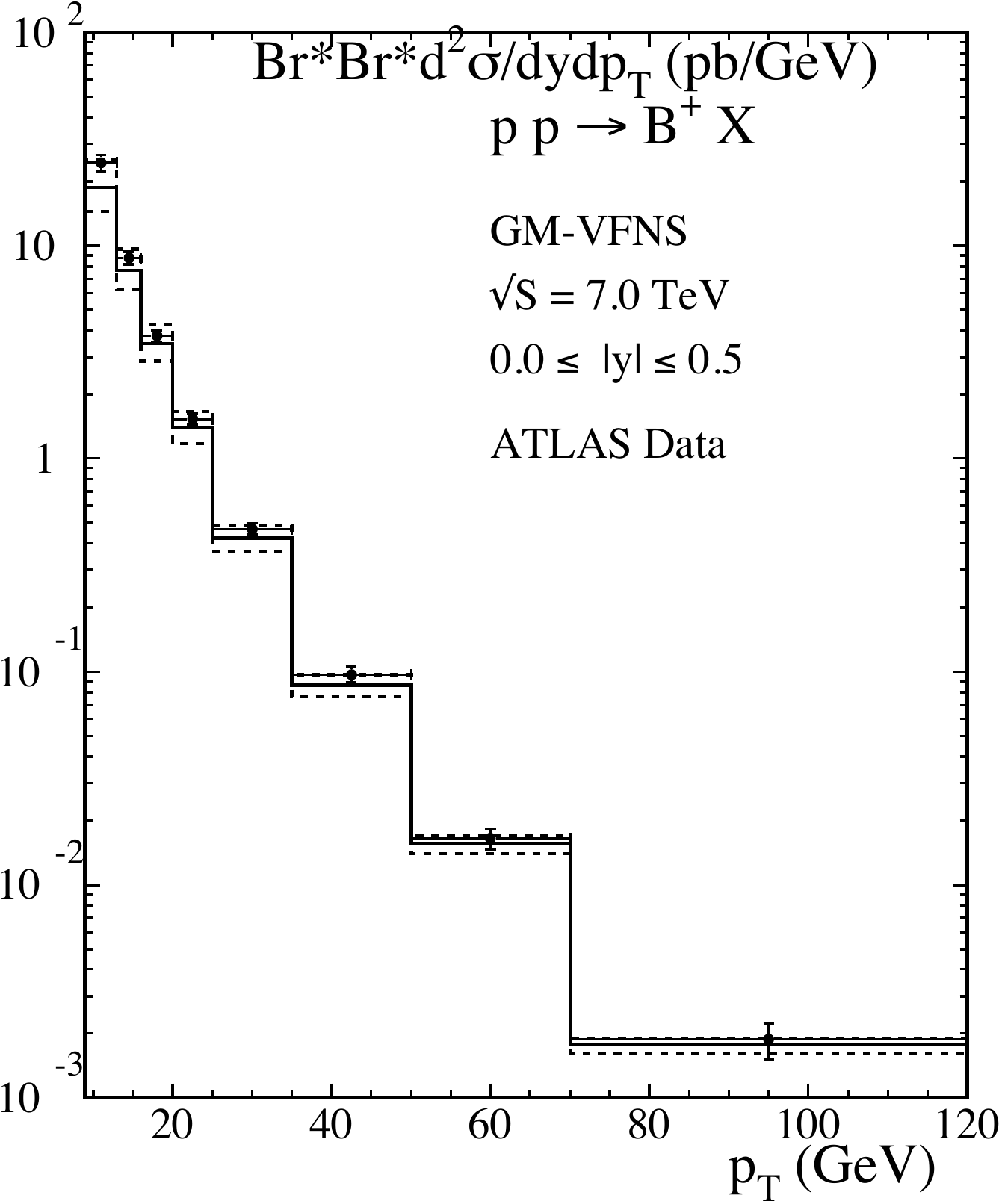}
\includegraphics[scale=0.55]{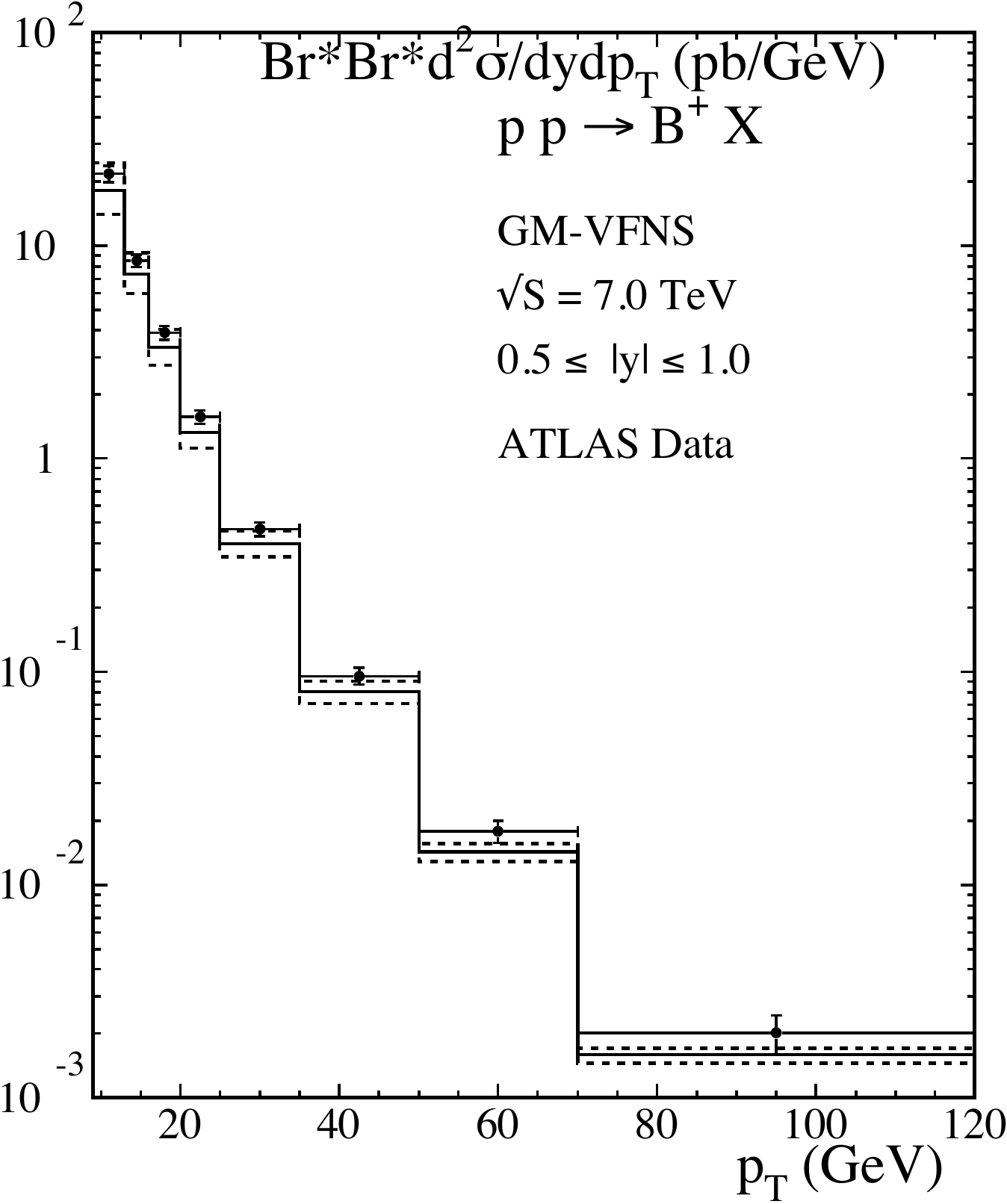}
\\
\includegraphics[scale=0.55]{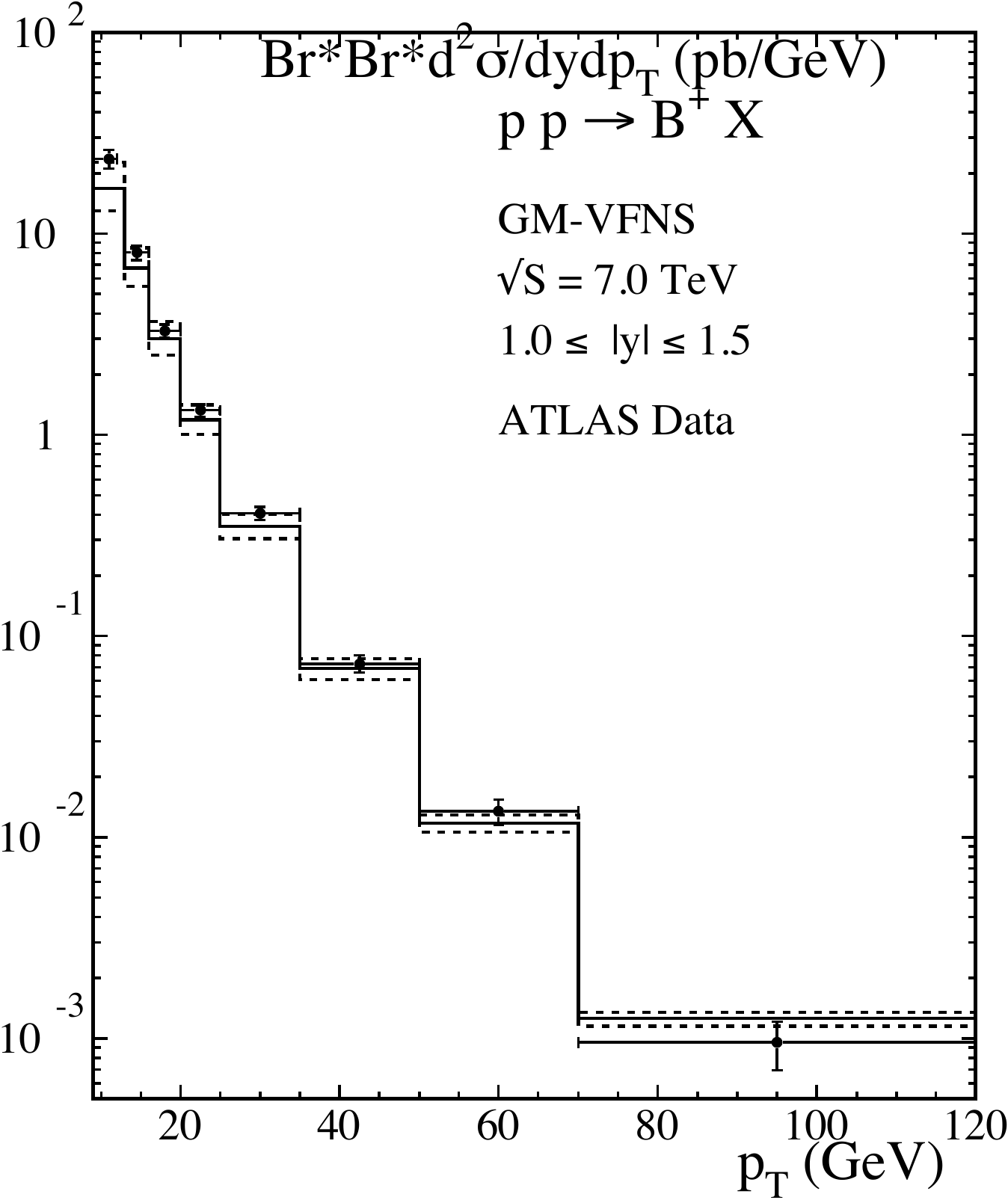}
\includegraphics[scale=0.55]{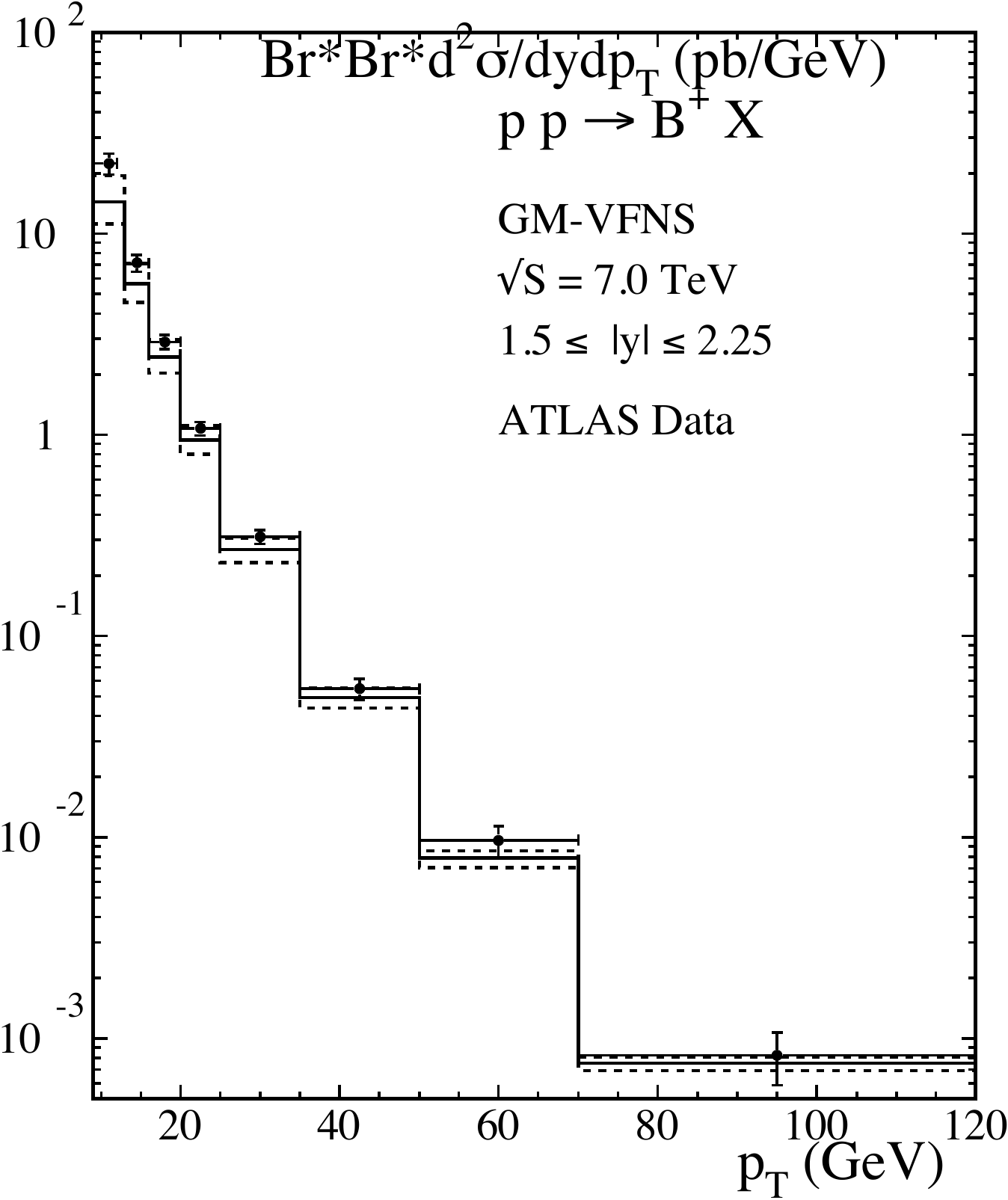}
\end{center}
\caption{
NLO predictions for $d\sigma/dp_T$ of $pp\to B^++X$ with $\sqrt{S}=7$~TeV and
$0\le y\le0.5$ (upper left panel),
$0.5\le y\le1$ (upper right panel),
$1\le y\le1.5$ (lower left panel), and
$1.5\le y\le2.25$ (lower right panel)
in the GM-VFNS are compared with ATLAS data \cite{ATLAS:2013cia}.
The default predictions and theoretical-error bands are evaluated as
in the right panel of Fig.~\ref{fig4ab}.
}
\label{fig8ad}
\end{figure}

\section{Conclusions}

Any theoretical prediction for hadronic collisions within perturbative QCD
requires the factorization of initial- and final-state 
singularities. This unavoidably introduces factorization 
scale parameters, which cannot be predicted from theory. Their 
choice must be based on physical arguments. We exploited 
the freedom in this choice to find a prescription that extends 
the reliability of NLO predictions in the GM-VFNS down to small $p_T$ values.
With scale parameters $\xi_{I} < 1$, 
it is possible to eliminate contributions from the heavy quark 
in the initial state. These contributions, dominated by the 
subprocess $gb \to gb$, are treated in the parton model 
with zero quark masses and would lead to an unphysical increase of 
the cross section for $p_T \to 0$. We showed, however, 
that with a judicious choice of $\xi_R$, $\xi_I$, and $\xi_F$, one 
can switch off these contributions in the small-$p_T$ range. 
Our prescription leads to a modified GM-VFNS yielding results 
in good agreement with recent data from the Tevatron and LHC 
experiments.
\bigskip

\section*{Acknowledgment}

We thank I.\ Bierenbaum for help in checking some of our
numerical calculations and J.\ Rojo for clarifying an
issue related to the heavy-quark threshold used in the NNPDF PDFs
\cite{Ball:2011mu,Ball:2011uy}.
This work was supported in part by the German Federal Ministry for Education
and Research BMBF through Grant No.\ 05H12GUE.

\clearpage



\end{document}